\documentclass[journal,article,accept,pdftex,moreauthors]{Definitions/mdpi} 
\firstpage{1} 
\pubvolume{1}
\issuenum{1}
\articlenumber{0}
\pubyear{2023}
\copyrightyear{2023}
\datereceived{} 
\dateaccepted{} 
\datepublished{} 
\hreflink{https://doi.org/} 

\Title{Effect  of transverse confinement on a quasi-one dimensional dipolar Bose gas}

\Author{Stefania De Palo $^{1,2}$, Edmond Orignac $^{3}$, Roberta Citro $^{4}$ and Luca Salasnich $^{5,6}$}

\address{$^{1}$CNR-IOM-Democritos, c/o SISSA (International School for Advanced Studies), Via Bonomea 265, I-34136, Trieste, Italy \\
$^{2}$Dipartimento di Fisica Teorica, Universit\`a Trieste, Strada Costiera 11, I-34014 Trieste, Italy\\
$^{3}$Univ Lyon, Ens de Lyon, CNRS, Laboratoire de Physique, F-69342 Lyon, France \\
$^{4}$Dipartimento di Fisica ``E. R. Caianiello'', Universit\`a degli Studi di Salerno and CNR-SPIN, Via Giovanni Paolo II, I-84084 Fisciano (Sa), Italy\\
$^{5}$Dipartimento di Fisica e Astronomia ``Galileo Galilei'', INFN and QTech, Universit\`a di Padova, via Marzolo 8, I-35131 Padova, Italy \\
$^{6}$INO-CNR, Unità di Sesto Fiorentino, via Nello Carrara 1, I-50019 Sesto Fiorentino (Firenze), Italy}

\abstract{We study a gas of bosonic dipolar atoms in the presence of a transverse harmonic trapping potential by using an improved variational Bethe ansatz, which includes the transverse width of the atomic cloud as a variational parameter. Our calculations show that the system behavior evolves 
from quasi-one dimensional to a strictly one-dimensional one by changing the atom-atom interaction, or the axial density, or the frequency of the transverse confinement. Quite remarkably, in the droplet phase induced by the attractive dipolar interaction the system becomes sub-one dimensional when the transverse width is smaller than the characteristic length of the transverse harmonic confinement.}


\DeclareMathOperator\erfc{erfc}
\begin{document}

\section{Introduction}
Trapped ultracold atomic gases offer a convenient and flexible platform to explore the fascinating aspects of many-body physics in one dimension \cite{cazalilla_review_bosons,mistadikis_1d}. 
In particular, in the last years the one dimensional dipolar Bose gas has been  largely investigated theoretically, see for instance Ref.~ \cite{baranov_condensed_2012}.
This kind of study is nowadays a vibrant topic of research, triggered by recent observations of self-bound droplets in attractive bosonic mixtures \cite{tarruell2018a,Semeghini2018,derrico2019} and in dipolar atoms \cite{pfau2016,luo2021}. In the real experiment, however, the system is not strictly one dimensional: the system made of identical atoms of mass $m$ is usually confined by a transverse harmonic potential of frequency $\omega_{\perp}$. Lately, we analyzed a more realistic case of finite transverse 
trapping frequency\cite{de_palo_variational_2020,depalo2021,baranov_condensed_2012} 
by using a variational Bethe ansatz of the ground state energy of a uniform dipolar gas in combination with a generalized Gross–Pitaevskii (GGP) equation. It has been found \cite{depalo2022} that the system evolves from a bright soliton-like into a droplet by increasing the atom number or the dipolar interaction strength.

 However, when the strength of the transverse trapping is reduced, the transverse width $l_{\perp}$ of the bosonic cloud could be quite different from the characteristic length $l_0=\sqrt{\hbar/(m\omega_{\perp})}$ of the transverse harmonic confinement. In the case of quasi-one-dimensional bosons with contact interaction, a generalized Lieb-Liniger approach, which takes into account effects of the transverse dynamics, was considered in Refs. \cite{salasnich2004,salasnich2005}.

{In this paper, extending the procedures of Refs. \cite{de_palo_variational_2020,depalo2021,depalo2022,salasnich2004,salasnich2005}, we analyze a quasi-one dimensional dipolar Bose gas by 
including, in an improved variational Bethe ansatz wavefunction, the effect of the transverse confinement.} In this way we study the evolution from a quasi-one dimensional dipolar bosonic system, where $l_{\perp} > l_0$, to an effectively one dimensional configuration, where $l_{\perp}\simeq l_0$.  We find that, by increasing the repulsive short-range interaction, the transverse width $l_{\perp}$ becomes larger and the effect is stronger for larger axial densities. On the contrary, when the dipolar attractive strength dominates, producing a droplet, the system becomes effectively one dimensional with $l_{\perp}$ very close to $l_0$ and, for large densities, it becomes sub-one dimensional with $l_{\perp}<l_0$. 

\section{Method: the variational approach for the energy functional}\label{sec:variational}

We start with the interacting dipolar gas of $N$ bosons in three-dimensions, {aligned in the $x-z$ plane by an external field along a direction $\hat{d}=\cos \theta \hat{x}+\sin\theta\hat{z}$, } in the presence of a transverse harmonic trap of frequency $\omega_\perp$, whose Hamiltonian reads
\begin{eqnarray}
H&=& \sum^N_i \left( -\frac{\hbar^2}{2 m} \nabla^2_i+ 
\frac{1}{2} m \omega^2_{\perp} (y^2_i+z^2_i)
 \right)
 + g_{3D} \sum_{i<j} \delta(|\vec{r}_i-\vec{r}_j|) \nonumber \\
 &&+ \sum_{i<j} \frac{\mu_0 \mu^2_d}{4 \pi |\vec{r}_i-\vec{r}_j|^3} \left(1 -3 \frac{[\hat{d}\cdot (\vec{r}_i-\vec{r}_j)]^2}{|\vec{r}_i-\vec{r}_j|^2}\right)
 \end{eqnarray}
where $m$ is the mass of the bosons, the contact strength is defined as
$g_{3D}=\hbar^2 \frac{4\pi a_{3D}}{m}$  via the  three-dimensional scattering length $a_{3D}$,
$\mu_0$ is the vacuum permeability and $\mu_d$ is the dipole moment.

\begin{figure}[H]
\begin{center}
\includegraphics[width=10.5cm]{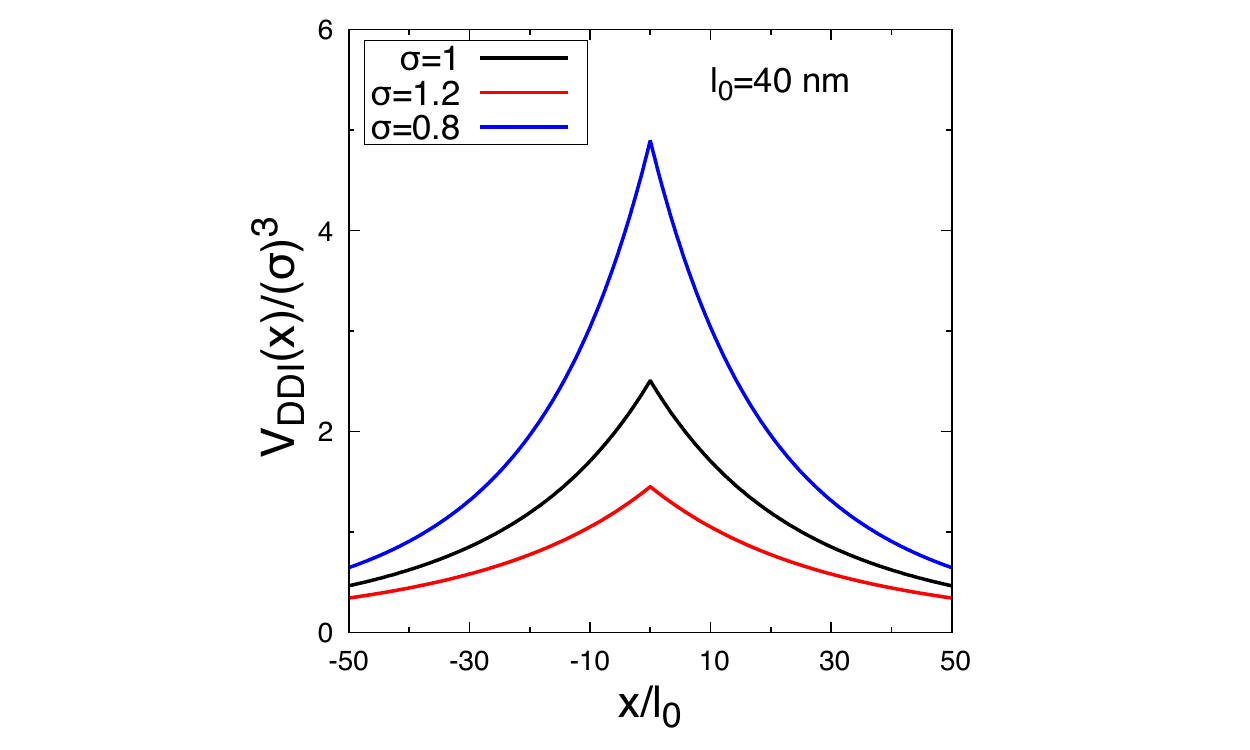}
\end{center}
\caption{For a fixed $l_0=40 nm$ we show how $V_{DDI}(x)/\sigma^3$ changes for
different values of $\sigma$, namely $\sigma=0.8,1$ and $1.2$.}
\label{fig:sigma_pot}
\end{figure}

The description by means of an effectively one-dimensional system relies on the assumption that
the trapping in the transverse direction is sufficiently tight to ensure that the gas
behaves as a one-dimensional gas. Here we wish to relax the assumption that the
transverse modes are frozen in the ground state of the trapping Hamiltonian and include
their effect in the trial  wavefunction ansatz \cite{salasnich2004,salasnich2005}:
\begin{eqnarray}
\phi(\vec{r}_1,\vec{r_2}\dots \vec{r}_N)=\psi_{1D}(x_1,x_2\dots,x_N) \Pi^{N}_{i=1}
e^{-\frac{z^2_i+y^2_i}{2 \sigma^2 l^2_0}}
\end{eqnarray}
where $\psi_{1D}$ is the wave function in the one-dimensional space while the wave function
in the transverse direction is modelled by the product of Gaussians
where $\sigma$ is the variational parameter that takes care of the spread of the density
in the transverse directions.
{When $\theta=0$, the dipoles are aligned along the $x$ axis, and the interaction has full rotational symmetry around that axis, thus justifying the choice of an isotropic $\sigma$. For $\theta \ne 0$, the model only has a reflection symmetry around the $xz$ and $xy$ planes, and a more general ansatz with a factor $\prod_i \exp[-\frac{z^2_i}{\sigma_z^2 l_0^2}-\frac{y^2_i}{2 \sigma_y^2 l^2_0}]$ could be considered. However,  in the presence of a tight harmonic trapping with full rotational symmetry, the isotropic ansatz is a good starting point.} Within this \textit{ansatz}, the projected one-dimensional Hamiltonian is:
\begin{eqnarray}
\nonumber
&& H_{Q1D}=-\frac{\hbar^2}{2 m} \sum_i \frac{\partial^2}{(\partial x_i)^2}
+ N \left(\frac{\hbar^2}{2 m l^2_\perp} +\frac{1}{2} m \omega^2_\perp l^2_\perp \right) \\
&&+\left[ g_{1D} -l_{\perp} \frac{8}{3} V(\theta,l_{\perp})\right]
\sum_{i<j} \delta(|x_i-x_j|) +V(\theta,l_\perp) \sum_{i< j} V_{DDI}(|x_i-x_j|/l_\perp)
\label{eq:hamiltonian_sigma}
\end{eqnarray}
with \cite{olshanii_cir,Santos_1,Reimann}
\begin{eqnarray}
&& l_\perp = \sigma l_0 \\
&& g_{1D}  =  \frac {g_{3D}}{2 \pi l^2_\perp} \\
&& a_{1D}  = -\frac{l^2_\perp}{a_{3D}} \\
&& V(\theta,l_\perp)= \frac{\mu_0 \mu^2_d}{4 \pi} \frac{1-3 \cos^2\theta}{4 l^3_\perp}.
\end{eqnarray}
\begin{eqnarray}
&&V^{1D}_{DDI} \left(\frac{x}{l_\perp}\right)=-2\left| \frac{x}{l_\perp}\right|
+\sqrt{2\pi} \left[ 1+ \left(\frac{x}{l_\perp}\right)^2 \right]
e^{\frac{x^2}{2 l_\perp^2}} \erfc \left[\left| \frac{x}{\sqrt{2} l_\perp} \right| \right]. 
\label{V1d_dd1}
\end{eqnarray}
In the new Hamiltonian (\ref{eq:hamiltonian_sigma}) the effective interaction {$V_{Q1D}(x)=V(\theta,l_\perp) V_{DDI}(x/l_\perp)$} depends explicitly on $\sigma$,  {at variance with previous derivations \cite{Santos_1,Reimann}},  and its effect can be appreciated in Fig.~\ref{fig:sigma_pot}: a tighter effective
 confinement ($\sigma<1$) leads to a stronger interaction than in the $\sigma=1$
case and viceversa.

Finally we estimate the variational ground-state energy of the system by using for $\psi_{1D}$
a Lieb-Liniger ground state wavefunction  with dimensionless interaction $\gamma$ as variational  parameter\cite{de_palo_variational_2020}. The trial energy to minimize is 
\begin{eqnarray}
\nonumber
&&\frac{E}{N}\left[ \frac{\hbar^2}{2 m} n^2\right]^{-1} = \epsilon(\gamma)-\left[ \gamma
-\frac{\gamma_0}{\sigma^2}\right] \frac{\partial \epsilon(\gamma)}{\partial \gamma}+2 \frac{a_d}{l_\perp} \frac{1-3 \cos^2 \theta}{n l_\perp}  \\
&& 
\left\{ 1+
\int^\infty_0 dq [S(\textit{q}; \gamma)-1]\left[ 
1-\frac{\textit{q}^2l^2_\perp}{2}e^{\textit{q}^2l^2_\perp/2}
\Gamma(0; \textit{q}^2 l^2_\perp/2)\right] \right\} + \frac{1}{(n l_0 \sigma)^2}+\frac{\sigma^2}{(n l_0)^2}
\label{eq:ene_sma}
\end{eqnarray}
with $\textit{q}= \pi n q $, $S(a;\gamma)$ the static structure factor\cite{cherny2008} {and $\Gamma[0,x]$ is the exponential integral function \cite{abramowitz_math_functions}}, 
{
\begin{equation}
\frac{\gamma_0}{\sigma^2}=\frac{2}{n}
\left\{-\frac{1}{a_{1D}} 
+ \frac{a_d}{l^2_\perp} \left[\frac{1-3 \cos^2\theta }{4}-\frac{8}{3}  \right]
\right\}
\end{equation}
}
and $a_d=m \mu_0 \mu^2_d/(8\pi)$and where $\epsilon(\gamma)$ is the ground state-energy of the Lieb-Liniger
model\cite{lang2017,ristivojevic2019,marino2019}. Using that \textit{ansatz} we have minimized the trial energy with respect to both $\gamma$ and $\sigma$ using standard minimization procedure\cite{brent_minimization_2002}.

\section{Results and Discussions}

\begin{figure}[H]
\includegraphics[width=10.5 cm]{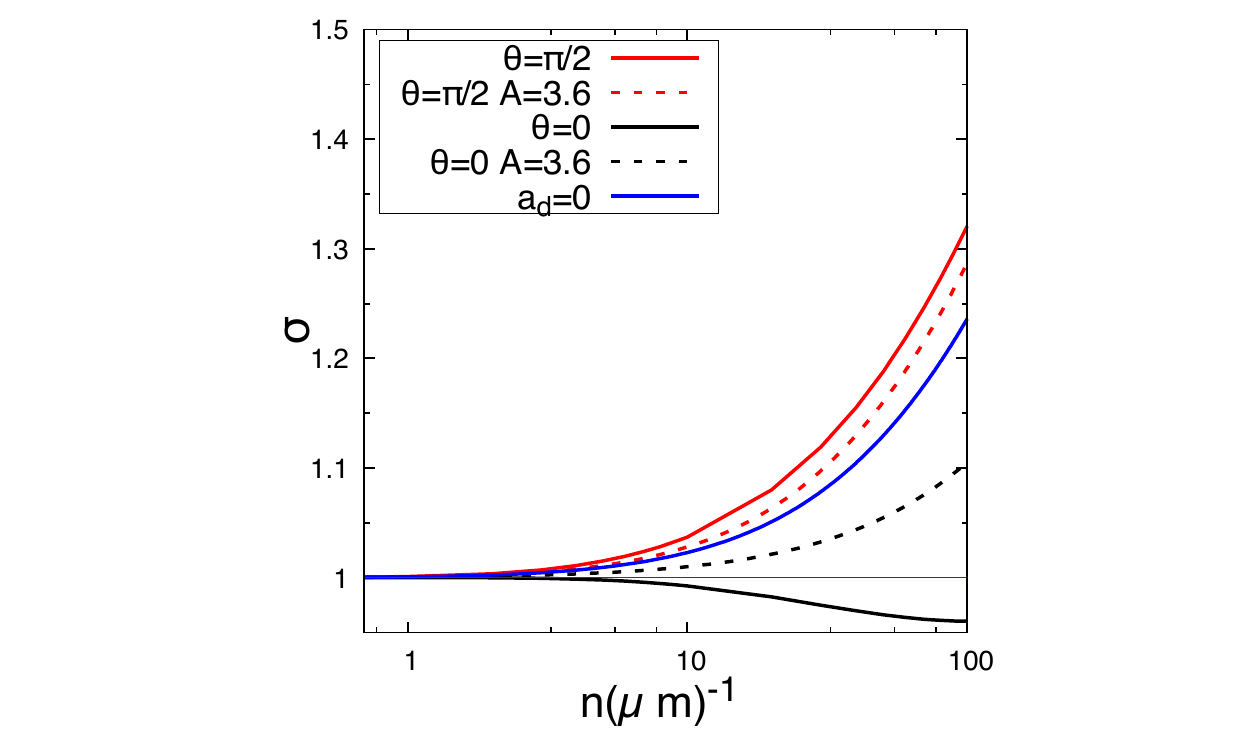}
\caption{Optimal values of $\sigma$ obtained with the different approximations as a function of density $n$. 
As a point of reference we show $\sigma$ for the
system without dipolar interaction, $a_d=0$ as a blue solid line.
The solid lines show the results obtained minimizing the energy functional (\ref{eq:ene_sma}), $\sigma_{sma}$,
while the dashed ones correspond to the minimization of the approximated Hamiltonian
(\ref{eq:hamiltonian_Avar_sigma}), $\sigma_A$. Red lines are for the repulsive case $\theta=\pi/2$
and black lines are for the attractive case $\theta=0$.
\label{fig:sigma_avar_cfr}}
\end{figure}

Before looking for the variational solution of the full single-mode energy functional (\ref{eq:ene_sma}) we consider a
first approximation where the short-range dipolar potential is replaced by an effective contact interaction
potential of strength $A=3.6$ \cite{Lev2018,de_palo_variational_2020,li_rapidity_2022}; within this approximation the Hamiltonian now reads:
\begin{eqnarray}
\nonumber
&& H_{Q1D}=-\frac{\hbar^2}{2 m} \sum_i \frac{\partial^2}{(\partial x_i)^2}  
+\left[ g_{1D}+l_{\perp} \left( A-\frac{8}{3}\right) V(\theta,l_{\perp})\right]
\sum_{i<j} \delta(|x_i-x_j|)  \\
&&+ \frac{N \hbar^2}{2 m l^2_\perp} +\frac{N}{2} m \omega^2_\perp l^2_\perp
\label{eq:hamiltonian_Avar_sigma}
\end{eqnarray}
Once we have dropped the short-range dipolar part, we are effectively back to the case already treated in 
\cite{salasnich2005}; the minimization of the energy functional with respect to $\sigma$ gives 
\begin{equation}
\frac{\partial E/N}{\partial\sigma}=\sigma^4+(n l_0)^2 \epsilon' \left[\frac{\gamma_0}{\sigma^2}\right]=0   
\end{equation}
where now the renormalized interaction is fixed by
{
\begin{equation}
\gamma_0= \frac{2}{n}
\left\{-\frac{1}{a_{1D}}+ \frac{a_d}{l^2_0} \left[\frac{1-3 \cos^2\theta }{4}+\left( A-\frac{8}{3} \right) \right]
\right\}
\end{equation}
}
In Fig.~\ref{fig:sigma_avar_cfr} we show, for a fixed scattering length
$a_{1D}/a_0=-8350$ ($a_0$ being the Bohr radius) and $l_0=57.3 nm$ and {$a_d=195 a_0$ as from Ref.~\cite{Lev2018}},  the optimal $\sigma$, i.e. the
rescaling parameter for the transverse confinement $l_\perp$. This is obtained by minimizing
the full Hamiltonian (\ref{eq:hamiltonian_sigma}) or minimizing Eq.
\ref{eq:hamiltonian_Avar_sigma} where we consider $A=3.6$, a constant independent from  the longitudinal density $n$ {as done in Ref.~\cite{Lev2018}}. 
We compare results for the repulsive case 
($\theta=\pi/2$, red lines) with the attractive case ($\theta=0$, black lines), 
together with a case without dipolar interaction ($a_d=0$, blue line).

Fig.~\ref{fig:sigma_avar_cfr} summarises the effect of the transverse confinement. Let's start with case 
with pure contact interaction, where, as expected upon increasing the density of the
system, the effective transverse width $ l_\perp=l_0 \sigma$ becomes larger due to the repulsive 
scattering between particles. When we add the repulsive dipolar interaction that effect becomes more
and more pronounced, with the optimal $\sigma_{sma} > \sigma_A$ in the whole range of densities.
The situation is different when we move to the attractive case, $\theta=0$ (we want to stress that for
given scattering length and $l_0$  the variational ground state energy \cite{de_palo_variational_2020} does have a minimum).
As shown in Fig.~\ref{fig:sigma_avar_cfr}, indeed, $\sigma_A>1$ while $\sigma_{sma}<1$, 
implying a reduction of the effective transverse width and therefore a tighter, more interacting system, which can be considered sub-one dimensional. We now devote the rest of the section to the quantitative discussion of the effect of including the transverse confinement in the variational approach.

\subsection{Repulsive dipolar interaction}

\begin{figure}[H]
\includegraphics[width=10.5 cm]{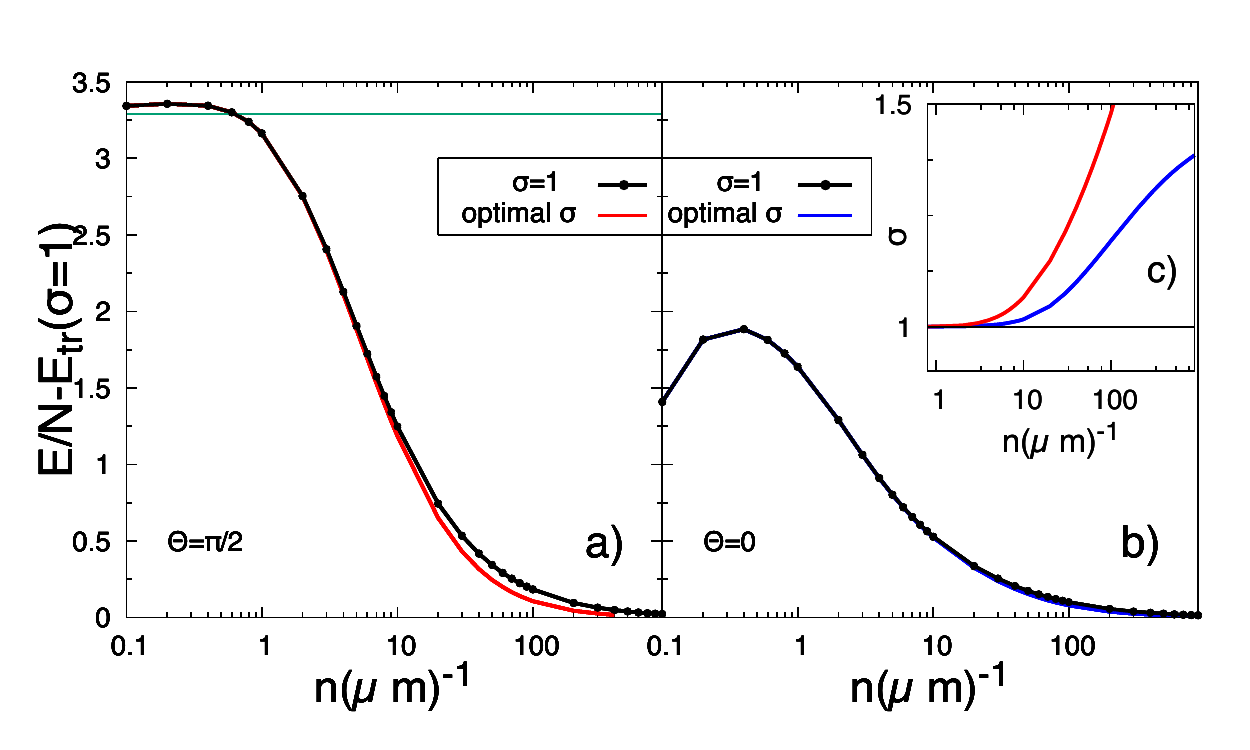}
\caption{ Ground state energy estimates, { in $\hbar^2/2m n^2$ units}, within variational ansatz 
for $l_0=57.3 nm$ and $a_{1D}/a_0=-2000$ as a function of particles density. The solid black lines are the energies obtained using $\sigma=1$ \cite{de_palo_variational_2020} while solid red lines are using 
the optimal $\sigma$ obtained from variational minimization. We show results for $\theta=\pi/2$
in panel (\textbf{a}) and $\theta=0$ in panel (\textbf{b}).{ We subtracted the transverse energy for $\sigma=1$ for clarity.}
In panel (\textbf{c}) we show the optimal values of $\sigma$ as a function of density for the cases 
reported in panel (\textbf{a}), solid red line,  and in panel (\textbf{b}), solid blue line. {The solid green line in panel a) is the 
limit of the energy in the Tonks-Girardeau limit.}
\label{fig:ene_cfr_2000}} 
\end{figure}

As already discussed, when the interaction between the particles is repulsive
the renormalized transverse confinement felt by the system $l_\perp$ is larger.
As expected this effect is more evident for loose confinement, 
in Fig.~\ref{fig:ene_cfr_2000} we show the equation of state, without the 
transverse energy contribution $E_{tr}/N= \frac{1}{(n l_\perp)^2} +\sigma^2 
\frac{1}{n l^2_0}$ for $\sigma=1$ as a function of density in two cases where the effective
interaction is repulsive.

For a quasi-one dimensional system like the experimental one  discussed in Ref. \cite{Lev2018}, in which 
typical values of the averaged density at the center of the trap range between $0.8 (\mathrm{\mu m})^{-1}$ and $3.1 (\mathrm{\mu m})^{-1}$, the effect of the inclusion of the transverse confinement 
is small with a the relative change of $\sigma \simeq 5\%$. This justified \emph{a posteriori} the approximation used in Ref. \cite{Lev2018}. 

\subsection{Attractive dipolar interaction: droplet region}

\begin{figure}[H]
\includegraphics[width=10.5 cm]{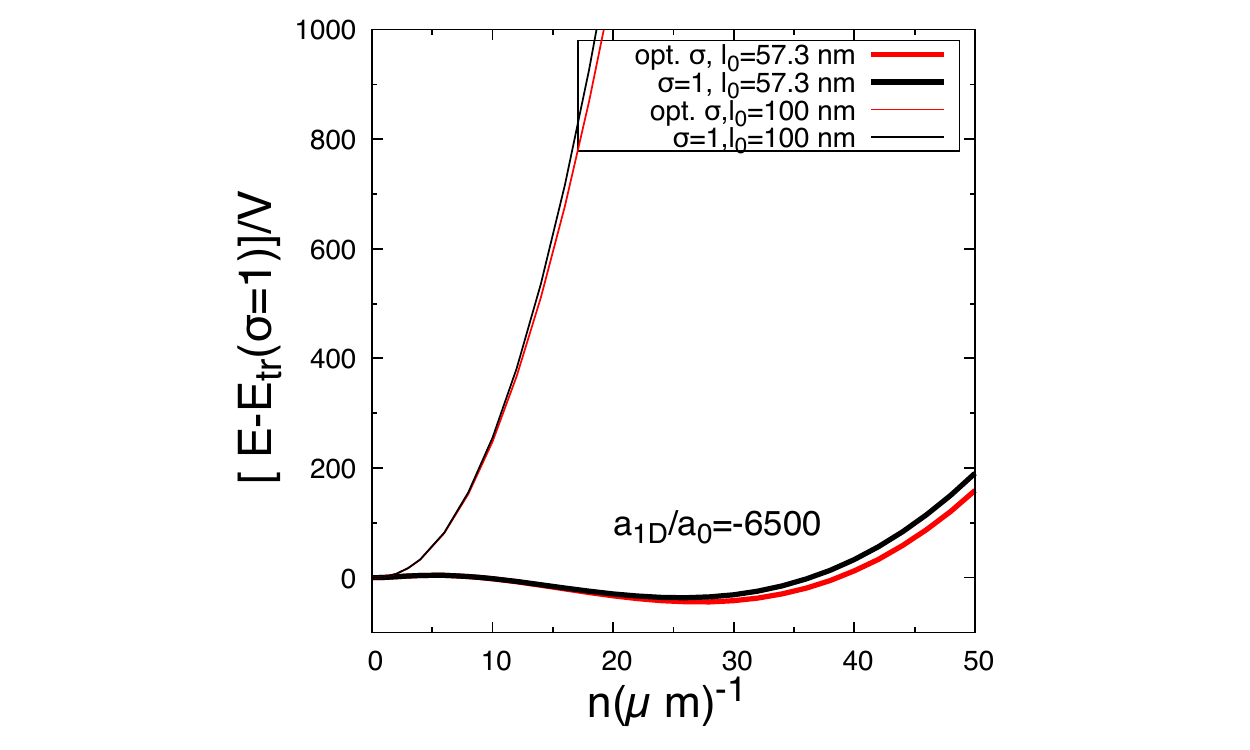}
\caption{Ground state energy estimates within variational ansatz for $a_{1D}/a_0=-6500$ for two selected values of $l_0$, namely $l_0=57.3 nm $ as from Ref.~\cite{Lev2018}
and $l_0=100 nm$. The solid black lines are the trial energies obtained using $\sigma=1$ while 
solid red lines are trial energies obtained using the optimal $\sigma$ from full minimization.  The thick solid lines are for
$l_0=57.3 nm$, while thin ones are for $l_0=100 nm$.
\label{fig:enev_6500_l0}} 
\end{figure}

When the attractive dipolar interaction becomes more relevant than the repulsive contact 
interaction in the variational Bethe-Ansatz \cite{de_palo_variational_2020} the estimated 
ground-state energy develops a deep minimum. Such feature favors the crossover from the gas
state towards the liquid-droplet state \cite{depalo2022}. The deep minimum occurs for quite
large densities and this, together with the variational character of the approach,  ensures 
lower energies when we add another variational parameter like $\sigma$ that governs the transverse confinement.
In Fig.~\ref{fig:enev_6500_l0} we assess the effect of varying the transverse trapping length $l_0$. 
As expected, reducing the trapping length enhances the interaction between particles and this enhances even more the variational parameter $\sigma$.

\begin{figure}[H]
\includegraphics[width=10.5 cm]{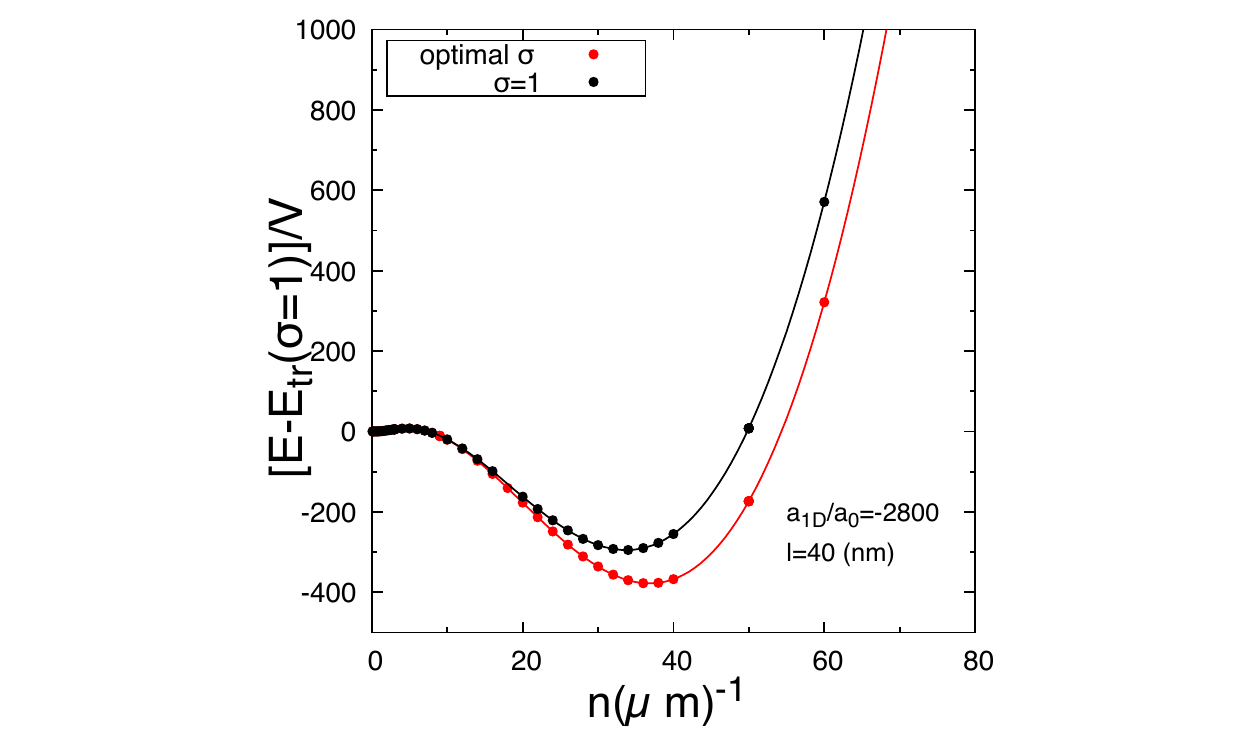}
\caption{Ground state energy within the variational ansatz for $l_0=40 nm$ and $a_{1D}/a_0=-2800$ as a function of particle density. The solid black lines are the 
trial energies computed using $\sigma=1$\cite{de_palo_variational_2020} while the solid red lines are computed with the optimal $\sigma$ obtained from variational minimization.
\label{fig:enev_2800_l40}} 
\end{figure}

In Fig.~\ref{fig:enev_2800_l40} we observe the effect of transverse confinement in a case where  the minimum
occurs at large densities and the lowering of the energy is sizeable. As a byproduct,  the minimum is also shifted to larger values of density. As a combined result the droplet is more stable and less sensitive to the longitudinal harmonic trapping. 

\section{Conclusions}

We used an improved variational Bethe ansatz approach to study a quasi-one dimensional dipolar gas taking care of the effect of transverse confinement. We considered an isotropic variational gaussian wavefunction that is a good reference choice for a tight confinement.  When the interaction
between particles is repulsive the effective transverse width of the cloud is increased and  the effect is visible for large densities. We found that for densities comparable with the ones of the experiments of Ref.~\cite{Lev2018} the trapping effect is negligible, thus justifying \emph{a posteriori} the assumptions of a pure one dimensional system ($\sigma=1$) in Ref. \cite{de_palo_variational_2020} {as well as the isotropic ansatz choice with $\sigma_x=\sigma_y$}. By contrast,  the formation of a droplet liquid state, that occurs when the attractive dipolar attraction prevails over the repulsive contact interaction could be more sensitive to the variation of the transverse trapping length.  The extended variational ansatz allows the system to become more tightly trapped in the transverse direction and denser in the longitudinal direction than in the strictly one dimensional case \cite{de_palo_variational_2020}. The net effect is a reinforcement of the stability of the droplet phase. 

{Our analysis could also be extended to the anisotropic case with gaussian wavefunction with $\sigma_z$, $\sigma_y$ spreading \cite{zinner2011} instead of a single variational parameter $\sigma$. This becomes relevant in multi-tube systems \cite{zinner2011}.} 
Beyond the static properties, this more accurate ground-state energy description could be used to assess the effect of transverse confinement on non-equilibrium properties, {\it e.g.} using a generalized Gross-Pitaevskii equation \cite{kolomeisky2000,dunjko_bosons1d,ohberg_dynamical_2002,oldziejewski_strongly_2019,depalo2022}. 

\acknowledgments{LS acknowledges for partial support INFN (iniziativa specifica "Quantum"), University of Padova (BIRD project "Ultracold atoms in curved geometries"), 
and the European Union-NextGenerationEU within the National Center for HPC, Big Data and Quantum Computing (Project No. CN00000013, CN1 Spoke 1: “Quantum Computing”).}

\conflictsofinterest{The authors declare no conflict of interest.}

\begin{adjustwidth}{-\extralength}{0cm}

\reftitle{References}

\end{adjustwidth}
\end{document}